\pdfoutput=1

\documentclass[%
 reprint,
 amsmath,amssymb,
 aps,floatfix
]{revtex4-2}

\usepackage[utf8]{inputenc}
\usepackage[T1]{fontenc}
\usepackage[x11names]{xcolor}
\usepackage{lineno}
\usepackage{amsmath}

\usepackage{graphicx}
\usepackage{dcolumn}
\usepackage{bm}
\usepackage{hyperref}

\begin{document}

\title{Morphological evolution via surface diffusion learned by convolutional, recurrent neural networks: extrapolation and prediction uncertainty 
}
\author{Daniele Lanzoni}
\affiliation{%
 Materials Science Department, University of Milano-Bicocca, Via R. Cozzi 55, I-20125 Milano, Italy
}%
\author{Marco Albani}
\affiliation{%
 Materials Science Department, University of Milano-Bicocca, Via R. Cozzi 55, I-20125 Milano, Italy
}%
\author{Roberto Bergamaschini}
\affiliation{%
 Materials Science Department, University of Milano-Bicocca, Via R. Cozzi 55, I-20125 Milano, Italy
}%
\author{Francesco Montalenti}
\affiliation{%
 Materials Science Department, University of Milano-Bicocca, Via R. Cozzi 55, I-20125 Milano, Italy
}%

\date{\today}

\begin{abstract}
We use a Convolutional Recurrent Neural Network approach to learn morphological evolution driven by surface diffusion. To this aim we first produce a training set using phase field simulations. Intentionally, we insert in such a set only relatively simple, isolated shapes. After proper data augmentation, training and validation, the model is shown to correctly predict also the evolution of previously unobserved morphologies and to have learned the correct scaling of the evolution time with size. Importantly, we quantify prediction uncertainties based on a bootstrap-aggregation procedure. The latter proved to be fundamental in pointing out high uncertainties when applying the model to more complex initial conditions (e.g. leading to splitting of high aspect-ratio individual structures). Automatic smart-augmentation of the training set and design of a hybrid simulation method are discussed.
\end{abstract}

\maketitle

\section{\label{sec:introduction}Introduction}

The intensive use of Machine Learning (ML) methods in recent years has proven a valuable tool in materials science and condensed matter physics~\cite{Butler2018Nature, Himanen2019AdvSci, Mehta2019PhysRep, Fiedler2021PRM, Lee2022PRM}. ML techniques have proven to be comparable or superior to traditional methods in tasks such as phase identification~\cite{Chung2022PRM}, experimental images processing~\cite{Yeom2021ActaMaterialia} and device property estimation~\cite{Zhu2022NatCom}. One of the areas that is attracting the most interest is the application of regression techniques to the development of interatomic potentials~\cite{Mueller2020JPC, Friederich2021NatMat}. The leverage of ML tools such as Neural Networks (NN)~\cite{Behler2007PRL, Sosso2012PRB,Behler2014JPCM, Behler2015IJQC} and Gaussian Processes~\cite{Bartok2010PRL, Bartok2015IJQC, Deringer2021ChemRev, Imbalzano2021PRM, Zeni2021NatCom} allowed for the construction of Force Fields far more accurate than classical semi-empirical potentials which however can be evaluated at a fraction of the computational effort of full ab-initio calculations~\cite{Zuo2020JPCA}. This computational speed-up makes it possible to tackle systems intractable by standard Density Functional Theory approaches~\cite{Deringer2018JPCL, Mocanu2018JPCB}.

Despite these improvements, the possibility of simulating complex mesoscale phenomena is still a hard task, both for the spatial and the time scales involved. In this respect, resorting to continuum scale methods is often the only solution.

The smoothing of discrete atomistic behavior in continuum fields typically leads to the formulation of evolution laws in the form of Partial Differential Equations (PDEs). It should nevertheless be remembered that even within continuum approaches the computational cost can be remarkable. For instance, fine meshes are sometimes required by Finite-Element-Method solvers, and/or stiff equations involving high-order derivatives might be involved (typical examples include Navier-Stokes equations in fluid dynamics~\cite{Glowinski1992AnnRevFluidMech} or degenerate Cahn-Hilliard equations in the presence of strong anisotropy~\cite{Salvalaglio2015CGD}). In the last few years, indeed, the interest in applying ML techniques for treating PDEs problems grew substantially~\cite{Kim2019CGF, Fulton2019CGF, Zhang2020CMAME, Teichert2019CMAME, MontesdeOcaZapiain2021npj, Yang2021Patterns}. Several of these new methods exploit the possibility of encoding spatial information of PDE solutions in images (when on 2D domains) or 3D volumetric data. This permits to leverage well-established ML techniques which have proven extremely successful in computer vision tasks. Among the different ML approaches that can be applied to predict image sequences, we find particularly intriguing the convolutional, recurrent neural network (CRNN) approach introduced in 2015 by Shi et al.~\cite{Shi2015NIPS} who modified the original long-short term memory (an instance of recurrent neural networks) replacing fully connected units with convolution operations. This variant, originally applied in the context of precipitation nowcasting, allows for pixel-wise prediction of the next image in a sequence, exploiting an efficient synergy of computer vision and recurrent NN approaches.

More recently, Young et al.~\cite{Yang2021Patterns} applied CRNNs (including the refinement introduced in Ref.~\cite{wang2018ICLR}) to some materials-science relevant phenomena, notably spinodal decomposition and dendritic growth as obtained from phase-field models. In this paper, instead, we apply CRNNs to another key problem in materials science, i.e. morphological evolution via surface diffusion~\cite{Mullins1957JAP, Li2009CommunComput}. Additionally, we implement a simple but effective scheme based on an ensemble model capable of providing an on-the-fly estimate of the prediction uncertainty. This allows one to understand under which conditions the model reliably handles extrapolation and when, instead, generated sequences are untrustworthy. With this respect, after building a convenient dataset based on phase-field simulations, we train, validate, and test the CRNN model on progressively harder tasks also involving heavy extrapolation.

The paper is organized as follows. In section~\ref{sec::phase-field} we present the phase-field model of surface diffusion and the selection criteria of the evolution produced to construct the datasets. Section~\ref{sec::ML} is devoted to the discussion of the specific implementation of our NN architectures, details the training procedure and presents how the prediction uncertainty estimation scheme is implemented. In section~\ref{sec::results}, we report performances on challenging settings, involving for example generalization to longer evolutions and more complex initial conditions. Finally, section~\ref{sec::conclusions} summarizes the current work and presents some of the possible future extensions.

\section{Data generation}
\label{sec::phase-field}

A simple curvature-driven surface diffusion model~\cite{MullinsJAP1957, LiCCP2009} is considered for the evolution of a two-dimensional profile $\Gamma$. Local material fluxes $J$ are set along the surface coordinate $s$ as a function of the surface gradients of the profile curvature $\kappa$ so that $\vec{J_s} = -M_s \nabla_s \kappa$, with $M_s$ a mobility constant. The resulting PDE problem is conveniently solved by a phase-field approach, tracking the profile implicitly by means of an order parameter $\varphi$, equal to 1 and 0 in the inner and outer domains delimited by the curve $\Gamma$ and with a diffuse interface along it, with finite, possibly small thickness $\epsilon$. By convention, the sharp profile is identified as the $\varphi$=0.5 isoline. The surface dynamics can then be rewritten in terms of the time-evolution of the phase-field itself, resulting in the Cahn-Hilliard equation set~\cite{LiCCP2009}:
\begin{equation}\label{eq::pf}
\begin{cases}
  \dfrac{\partial\varphi}{\partial t} & =\nabla\cdot M(\varphi)\nabla \kappa \\
 g(\varphi)\kappa & =-\epsilon\nabla^2\varphi+\frac{1}{\epsilon}W'(\varphi)
\end{cases}
\end{equation}
with $W(\phi)=18\varphi^2(1-\varphi)^2$ a double-well potential with minima in the two "bulk phases" $\varphi$=0 and 1 and $g(\varphi)=(5/3)W(\varphi)$ a stabilizing function introduced for numerical convenience~\cite{GugenbergerPRE2008,SalvalaglioMMAS2021a,SalvalaglioMMAS2021b}. The degenerate mobility function $M(\varphi)=M_s(36/\epsilon)\varphi^2(1-\varphi)^2$ is used to restrict the dynamics within the interface region, thus returning the proper surface diffusion behavior in the sharp-interface limit for $\epsilon\rightarrow 0$.

\begin{figure}
    \includegraphics[width=\linewidth]{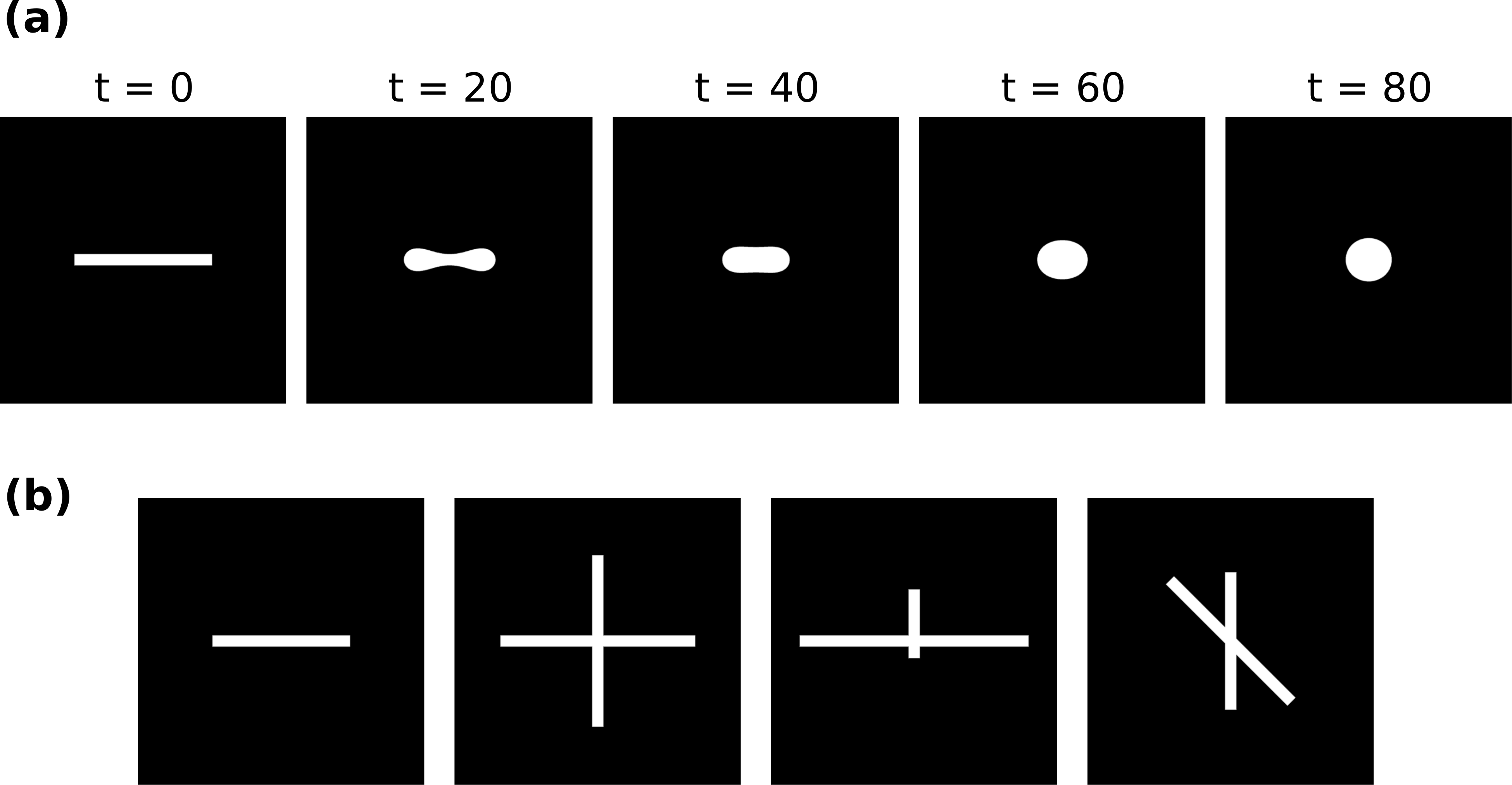}
    \caption{\label{fig::figure1} (a) Prototypical evolution of a rectangular domain by isotropic surface diffusion. Intermediate steps present a characteristic hourglass shape, while the final state is circular. (b) Examples of shapes present in the training set (from left to right: simple bars, symmetrical crosses, asymmetrical crosses and 45° crosses).}
\end{figure}

The numerical solution of the PDE system~\eqref{eq::pf} is performed by Finite-Element Method, using the Adaptive Multidimensional Simulations (AMDiS)~\cite{VeyCVS2007,WitkowskiACM2015} toolbox. Both a semi-implicit time-integration scheme and adaptive local mesh refinement are exploited for numerical efficiency. The lenght scale in the simulations has been scaled with respect to the interface thickness $\epsilon$. A 400$\times$400 domain is considered to embed all test geometries with finest mesh resolution of $\sim 1/7$ to properly represent the diffuse interface. Time is scaled with respect to the surface mobility coefficient $M_s$. A fixed time step of $10^{-5}$ is set and a total evolution time of $0.6$ is considered for all simulations, as sufficient to reach a convex shape for all initial conditions considered. The choice of a fixed integration time step over more sophisticated adaptive schemes has been chosen since the CRNN structure considers sequence elements as equally spaced in time. This parameter set has been tested to ensure numerical convergence and stability for all geometries considered in the present study, offering the best tradeoff between accuracy and execution speed, the latter being important to produce a sufficiently large dataset for the ML training. As an example, the evolution of a simple rectangular domain is shown in Fig.~\ref{fig::figure1}(a).

For the present study, the dataset was constructed considering only prototypical shapes which could be obtained by rectangles and simple intersection of rectangles: simple bars, symmetrical crosses, asymmetrical crosses and 45° crosses. An example for all possible initial topologies is reported in Fig.~\ref{fig::figure1}(b). Different dataset elements were constructed by scanning rectangles aspect ratios between 1 and 20, keeping the short side at 8 units of $\epsilon$. The limitation in the aspect ratios prevents domain splitting due to Rayleigh-like instability~\cite{Salvalaglio2018APL}. All the constraints in the dataset construction were intentional, as one of the objectives of our work was to understand to which extent a Machine Learning algorithm is able to generalize to unobserved configurations given a simple and controlled set of examples.

\section{Machine learning methods}
\label{sec::ML}
\subsection{Convolutional Recurrent Neural Network}
\label{sec::convRNN}

As stated in the introduction, PDEs in 2D can be conveniently encoded as images. This approach is particularly straightforward in the case of phase field models, since the value of the order parameter $\varphi$ can be restricted between 0 and 1. Successive states of the morphological evolution of an initial shape under surface diffusion, therefore, can be represented by sequences of grayscale images.

In recent years, convolutional neural networks (CNN) have proven to be one of the most effective machine learning architectures for tasks involving data with spatial structures. Invented in the late eighties \cite{LeCun1989NeurComp, LeCun1989IEEE}, they are at present one of the standard tools in tasks such as audio data analysis \cite{Dieleman2014IEEE, Kiranyaz2021MSSP}, image classification \cite{Li2021IEEE, Rawat2017NeurComp} and inference from volumetric data \cite{Milletari2016IEEE, Maturana2015IEEE}. The main building block of CNNs consists in the convolution operation. Starting from raw data, hierarchically more complex "abstract" features may be extracted by successive application of convolutional kernels (typically in the form of square matrices) containing learnable parameters. Recently, convolutional structures revealed their potential also in condensed matter and materials science research. For instance, they have proven effective in analyzing experimental images \cite{Yeom2021ActaMaterialia}, in accelerating or reducing the dimensionality of complex continuum model simulations \cite{Kim2019CGF, Fulton2019CGF, Zhang2020CMAME, Teichert2019CMAME} and in tackling mesoscale simulations \cite{Yang2021Patterns}.

The CNNs' capability of identifying spatial correlations is effective in tasks involving PDEs because they encode several mathematical properties by design~\cite{goodfellow2016BOOK, aggarwal2018BOOK}. First, convolutions are local operations: pixels which are farther apart than the kernel size are not allowed to "communicate". If CNNs are used to approximate PDEs, this directly translates in the locality of (partial) differential operators. In the present case, for instance, the isotropic Cahn-Hilliard equation predicting the evolution of a domain only depends on its local curvature.

A second important property of convolution transformations is that they are equivariant with respect to the group of spatial translations. This is beneficial to the training procedure, both in terms of training time and quality of the final result, as the NN does not need to learn the underlying symmetry group. Correspondingly, predicted evolution will not depend on the absolute position of the initial shape by construction. The use of convolution operators also allows for a straightforward implementation of periodic boundary conditions, which can be simply obtained by using a so-called "circular padding" \cite{Schubert2019IEEE, Kayhan2020CVF}.

Last, if all transformations in the NN are either convolutions or pixel-wise (fully convolutional NNs), the model can operate on images of arbitrary size. In practice, this means that the CNN can be trained on images representing small simulation cells and, once parameters are optimized, the learned evolution rules can be applied on much bigger domains, considering the pixel dimension as fixed~\cite{goodfellow2016BOOK, aggarwal2018BOOK}. This allows one to tackle mesoscale systems with simulation cells several time the size of the original dataset, an advantage clearly demonstrated in~\cite{Yang2021Patterns}.

While CNNs are suited for dealing with local spatial correlations, the goal of the current work is to approximate morphological time evolution. This kind of task involves analyzing and generating temporal sequences. In this field, RNNs emerged as one of the most effective machine learning strategies. Recurrent structures usually exploit a hidden state, which can be considered as a sort of "memory" of previous states of the system and is capable of capturing time correlations \cite{Lipton2015arXiv}. There are several possible implementations for RNNs, but in the last years Long-Short Term Memory (LSTM,~\cite{Hochreiter1997NeurComp}) and Gated Recurrent Units (GRU,~\cite{Cho2014SSST}) have emerged as the most used and effective. In our case, we opted for the latter, as GRUs often offer performances comparable to LSTMs with a lower number of parameters~\cite{Chung20141arXiv}, which translates to faster evaluation and a reduced possibility of overfitting.

Similarly to fully convolutional NNs, which can transform images of any size, RNNs can elaborate sequences of arbitrary length. This means that a RNN based model can use a partial evolution of any length (possibly a single initial state) as an input and then generate sequences of arbitrary length as well \cite{goodfellow2016BOOK, aggarwal2018BOOK}.

As the objective of the current work is approximating the time evolution of systems described by PDEs, the spatial features learning capabilities of CNNs and the temporal ones of RNNs should be used simultaneously. Following the work of \cite{Shi2015NIPS} and \cite{Ballas2016ICLR}, we implemented a Convolutional GRU (ConvGRU) cell as the building block of our NN structure. The main idea consists in having a hidden state which is itself structured as a (multiple channel) image and in replacing linear transformations of regular GRUs with convolutions. The result is a neural network capable of learning spatiotemporal correlations on a pixel level. Notice that our approach, while similar to the one of \cite{Yang2021Patterns}, does not require convolutions involving temporal variables, thus using a smaller number of parameters. Additionally, we note that, from a purely technical point of view, extension of the present approach to three spatial dimensions is straightforward, but the actual problem stems in GPU requirements and in the effort needed to generate the dataset. 

Here, for the sake of reproducibility, we list all hyperparameters. The paragraph can be skipped by uninterested readers. The overall CRNN structure used in the current work is composed of two stacked ConvGRU cells, for a total of $\sim 2.8\times 10^5$ parameters. Both hidden states have $35$ channels, convolution kernels are set to $5\times5$, stride $1$ (i.e. the convolution runs over all pixels), and circular padding encoding periodic boundary conditions by construction are used. No pooling operations were exploited. The hidden layer is converted to the next state by a $3 \times 3$ convolution with no pooling, stride $1$ and circular padding followed by a sigmoid activation function reducing the output image in the range $[0,1]$. This set of hyperparameters produced the most satisfying results while containing computational costs in our tests. 

We conclude this section by emphasizing the fact that all computational operations involved in a trained CRNN are typically much cheaper computationally than a FEM solver (by orders of magnitude for all examples discussed in the paper). Additionally, they can easily be implemented on parallel machines (or GPUs). Lastly, the timesteps between one state and the following in the evolution obtained by machine learning is much longer than the one used in traditional schemes. All these advantages contribute in compressing the computational times of the predicted evolution, allowing to tackle configurations which would be unconvenient to simulate by traditional methods (e.g. long time evolutions, large or complex systems).

\subsection{Equation symmetries and training}
\label{sec::symmetries}

As specified in the previous Section, one of the main advantages of CRNN is the encapsulation of some symmetries of the underlying equation, such as translational equivariance and locality. On the other hand, standard convolutions do not encode other symmetries which are nonetheless present in the degenerate Cahn-Hilliard equation. In order to leverage this additional features, we employed two strategies: modification of the loss function and data augmentation.

In supervised machine learning tasks, the objective of the training procedure consists in the minimization of a loss function $L$, which can be considered as the deviation between the predicted output and the correct one present in the training set. The base for the loss function used in this work was the commonly used in regression tasks Mean Squared Error Loss:
\begin{equation}
    L(\theta) = \frac{1}{N_{TS} T} \sum_{i=1}^{N_{TS}} \sum_{t=1}^T \langle (\varphi_i(t) - \hat{\varphi}_i(t | \theta))^2 \rangle_x
\end{equation}
here, $\theta$ represents the set of NN parameters, $i$ indexes the training set elements, $t$ indexes the timestep and runs from 1 to the total length of the sequence $T$, $\varphi$ is the true sequence of the system temporal evolution in the training set, $\hat{\varphi}$ is the time evolution as predicted by the CRNN and $\langle . \rangle_x$ denotes spatial average.

In order to enforce global conservation of the order parameter $\varphi$ (as dictated by Cahn-Hilliard equation), the original loss function has been modified in order to penalize non-conserving dynamics:
\begin{eqnarray}
    \label{eq::loss_function}
    \widetilde{L}
    (\theta) = \frac{1}{N_{TS} T} \sum_{i=1}^{N_{TS}} \sum_{t=1}^T \langle (\varphi_i(t) - \hat{\varphi}_i(t|\theta))^2 \rangle_x + \\ \alpha (\langle \varphi_i(t) \rangle_x - \langle \hat{\varphi}_i(t|\theta) \rangle_x)^2
\end{eqnarray}
where $\alpha$ sets the penalization strength. In our tests, we observed that a value of $\alpha=2$ led to the best performances.

Another symmetry to be taken into account is the equivariance of the evolution with respect to rotations of the initial condition. In order to enforce it, we adopted a data augmentation strategy, in which every time sequence is rotated by a random angle before being passed to the neural network. This approach regularizes parameters in convolutional kernels, forcing them to approximately respect rotational equivariance.

A last data augmentation procedure has been used to encode in the Neural Network the symmetry in the double well potential and mobility function with respect to the transformation $\varphi \rightarrow 1-\varphi$ (see Sec.~\ref{sec::phase-field}). This means that the negative version of evolution in Fig.~\ref{fig::figure1}(a) is still a valid solution to the Cahn-Hilliard equation. The physical interpretation of this is that holes in a material matrix will undergo the same morphological evolution as a material domain with an identical initial shape. Being a discrete symmetry, it is possible to expose the CRNN to all sequences both in the positive ($\varphi$) and negative ($1-\varphi$) version. This way, gradients in the backpropagation algorithm are never computed on positive or negative only variants, reducing biases with respect to this symmetry.

\subsection{Prediction uncertainty estimation}
\label{sec::pred_uncertainty}

One of the main pitfalls that can be encountered during the training of NNs and other machine learning models is that it is not clear if the produced answer is reliable or the model is being used in extrapolation conditions~\cite{Musil2019JCTC}. This problem is particularly critical in the current case, as the training set was restricted on purpose to simple initial conditions, in order to avoid evolution which require computationally intensive simulations and specialized algorithms (see discussion in Sec.~\ref{sec::results}). In general, a prediction uncertainty quantification is essential to detect such issues. A common strategy to asses this problem is through testing: model performances are monitored on an ad-hoc test set, representing typical tasks the machine learning model will encounter once deployed~\cite{goodfellow2016BOOK, aggarwal2018BOOK}. In our case, however, the construction of a proper test set presents several challenges. For example, it is not clear how a representative set of initial conditions should be built. Furthermore, such set should comprise also configurations presenting complex interactions which, as it was already mentioned, are inconvenient to simulate in high number because of computational effort.

An alternative solution is leveraging a resampling procedure known as bootstrap aggregation (bagging) \cite{Breiman1996ML}. This technique allows for the construction of an ensemble of models, whose prediction capabilities are higher than the individual models' one. In addition, the dispersion of the ensemble predictions represents an estimation of the prediction uncertainty \cite{Tibshirani1996neco}.

In our case, we opted for a simple implementation of bagging. Starting from the original training set composed of $N$ sequences, a number of new sets with the same size are generated by random extraction with replacement. These new sets are called bootstrap samples (BS). Extraction with replacement guarantees that BSs, though drawn from the same original set, are different. For each BS an independent model is fitted, thus generating the neural network ensemble. It can be proven from a statistical point of view that, if the learned models present an unstable enough training procedure, the aggregate prediction of the ensemble is more robust and reliable than individual ones \cite{efron1992COLLECTION}. Another key advantage is that variance and/or standard deviation between individual predictions of the models in the ensemble can be calculated. In the current framework, this allows for the definition of an on-the-fly, pixel-wise prediction uncertainty, thus the identification of regions and instants in which the predicted evolution is unreliable.

To form the aggregate prediction, we used simple mean. On a qualitative level, when models inside the ensemble provide diverging predictions, aggregation of the individual predictions will produce a gray "halo", as the mean of the predicted values of $\varphi$ is different from 0 or 1. At the same time, the ensemble standard deviation allows for the definition of a confidence interval on a quantitative level.

In order to maximize the independence in the training procedure, bootstrapping has been applied on both training ($\sim 1200$ sequences) and validation ($\sim 250$ sequences) sets and an ensemble of 15 models was constructed. This value was chosen as a reasonable compromise between the ensemble size, training times and prediction computational costs, the latter two scaling linearly with the number of models.

\subsection{Training, validation and testing}
\label{sec::train_valid_test}
As anticipated, training has been performed with the modified Mean Squared Error Loss $\widetilde{L}$ reported in equation~\eqref{eq::loss_function}. NNs' parameters were optimized using the standard Adam algorithm \cite{Kingma2015arXiv}, as implemented in the Pytorch library~\cite{NEURIPS2019_9015}.

Training and validation sets are composed by sequences of 150 snapshots, starting 0, 25 and 50 timesteps from the initial state of the system. This choice allows for the exposure of the models to different initial conditions. The time difference between each timestep is 250 time integration steps of the high-fidelity phase-field FEM solver (see Sec.~\ref{sec::phase-field}). Sequences originating from the same initial condition were exclusively inserted in either the training or the validation set, in order to remove correlations which would affect validation loss.

\begin{figure*}
    \includegraphics[width=\linewidth]{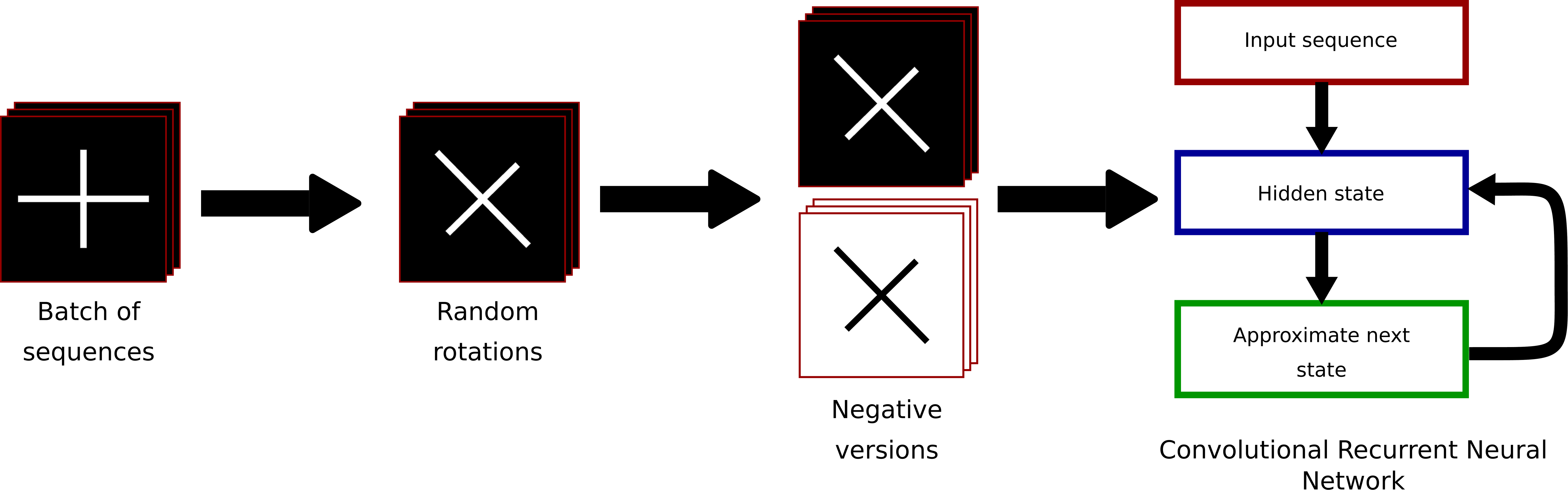}
    \caption{\label{fig::figure2} Scheme of the training procedure. A single parameters optimization step consist in a random extraction of a batch of sequences which are converted to the right resolution size, randomly rotated and passed to the CRNN together with their negative versions.}
\end{figure*}

From a practical standpoint, training is summarized in Fig.~\ref{fig::figure2} and proceeds as follows. A batch of sequences is extracted from the training set, converted to a $90 \times 90$ pixels format and randomly rotated. The batch is then augmented with the negative version of the evolution through the transformation $\varphi \rightarrow 1-\varphi$. Next, truncated subsequences are passed to the Convolutional Recurrent structure, whose task is to generate its completion. The truncation point is selected at random for every batch. This choice allows for a compromise between the full, more demanding task of generating a sequence starting from a single initial state and the easier (but far from the actual application mode) task of completing a partial sequence. The produced sequence is then compared with the one in the training set, loss function is evaluated and parameters are updated.

\begin{figure*}
    \includegraphics[width=\linewidth]{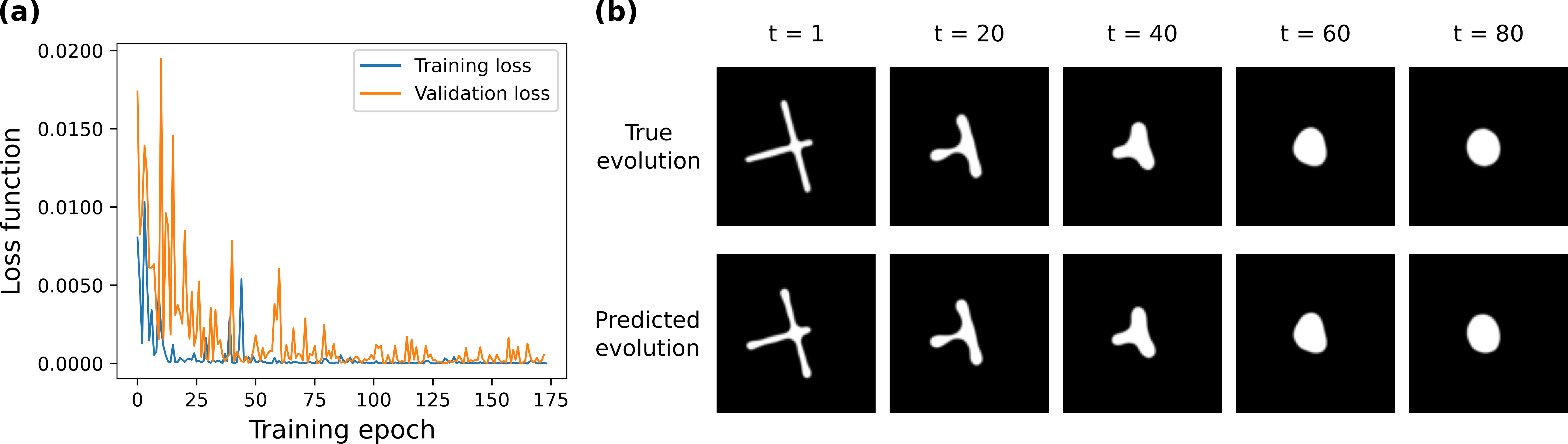}
    \caption{\label{fig::figure3} (a) Training and validation loss value as a function of the epoch number for one model in the ensemble. (b) Example predicted evolution for a validation set element and the corresponding high-fidelity FEM sequence.}
\end{figure*}

A training plot is reported for an individual model in the bootstrap aggregate in Fig.~\ref{fig::figure3}(a). Validation loss is always calculated by providing the model only the initial state of the evolution, as this is the actual task the network is required to tackle with new, unobserved initial conditions. Although noisy, after $\sim 75$ epochs (in one epoch all elements in the training set are presented once to the NN model) the validation loss settles, indicating the training procedure has converged. No increasing trends can be observed, confirming the absence of overfitting. Oscillations in the loss value as a function of epochs are induced by the randomness of the optimization procedure. The evolution of a validation initial condition as predicted by the model is reported in figure Fig.~\ref{fig::figure3}(b), confirming the model parameters have converged.

\begin{figure}
    \includegraphics[width=\linewidth]{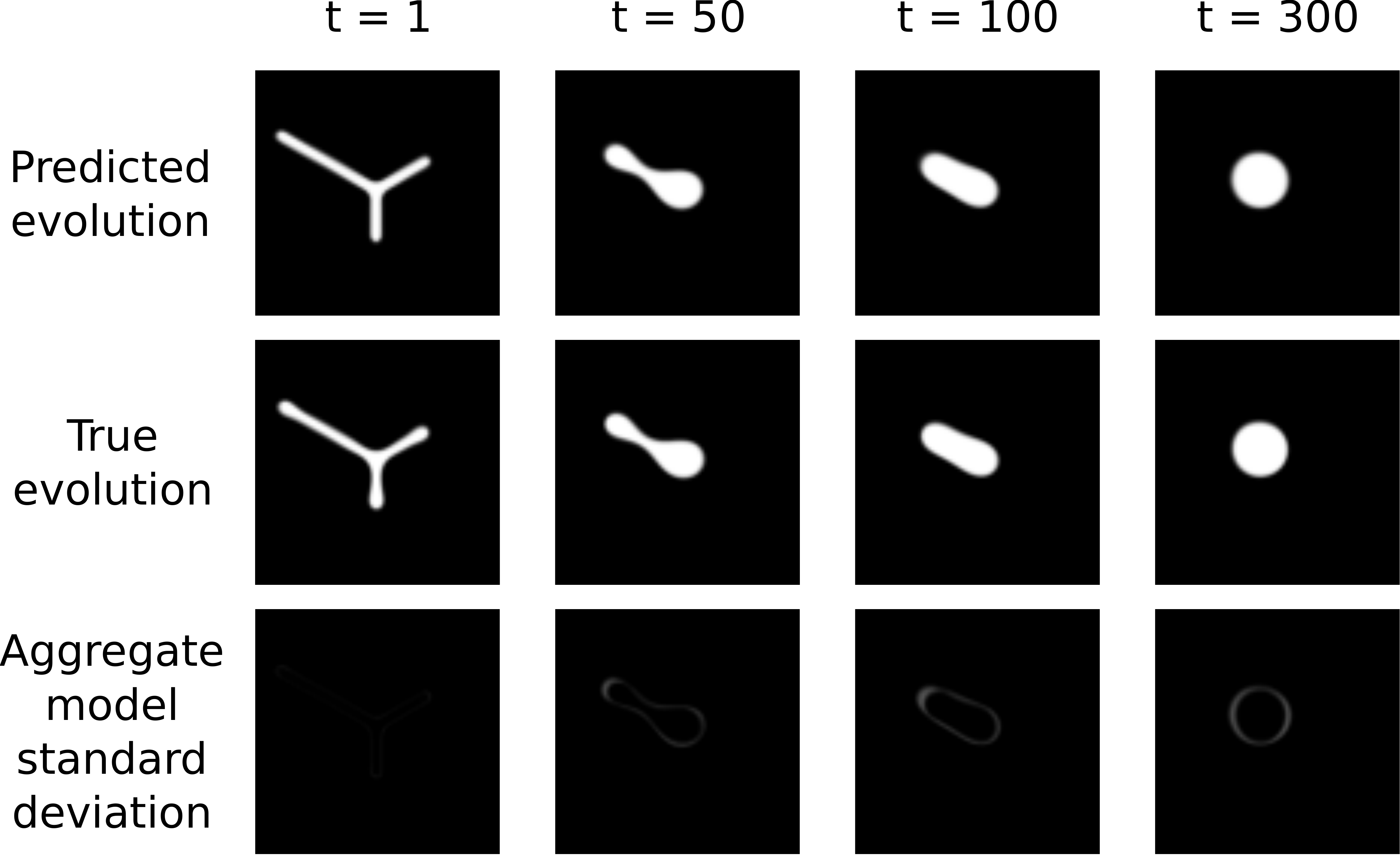}
    \caption{\label{fig::figure4} Example evolution of a Y shape as a testing of the aggregate model performances (not included in training). The lack of high model variance regions confirms that this example is far from extrapolation conditions. Generated sequences are twice as long as the ones of the training procedure.}
\end{figure}

As an additional confirmation of the training procedure results, we inspected evolution predicted starting from previously unobserved and qualitatively different Y-shapes. Arms of these new initial conditions have the same thickness as rectangles and crosses present in the original dataset. Inspection of the example evolution reported in Fig.~\ref{fig::figure4} reveals that predicted evolution exhibits only negligible deviations from the true one. At the same time, prediction uncertainty is low, confirming that the aggregate model is not in extrapolation conditions even though the total number of frames predicted is 300, double the 150 used during training, suggesting good generalization capabilities to long time evolutions. We emphasize that the close agreement on this specific test configurations cannot be considered as an indication of performances on general initial conditions, as a single class of shapes has been tested. Our concern, here, is instead to check the ensemble model behavior for progressively harder tasks.

\section{Results and Discussion}
\label{sec::results}
In the following section, we test the capabilities of the presented machine learning approach in predicting long term morphological evolution (i.e. for times much longer than the 150 time steps used in training) and in capturing important scale-dependent behaviors which were not explicitly inserted in the training set. We conclude the section by inspecting generalization capabilities and limitations in the case of complex initial conditions with many interacting shapes.

\subsection{Stationary state stability}
One of the main concerns that should be addressed is the stability of the NN prediction for long time evolution. While a simple test was provided with the evolution of Fig.~\ref{fig::figure4}, we perform now a more in depth analysis which also considers an unobserved topology. In particular, the time evolution of an annular shape stationary state configuration is considered. The goal of this test is threefold: the first check consists in verifying that the aggregate model correctly predicts that the hole in the center of the domain is a stable configuration (as the curvature and, thus, the chemical potential is uniform); second, we check if the stability is predicted for times exceeding 150 time steps, indicating good time generalization capabilities of the model; third, we verify that prediction uncertainty rises if and as soon as the generated evolution diverges from the correct one.

\begin{figure}
    \includegraphics[width=\linewidth]{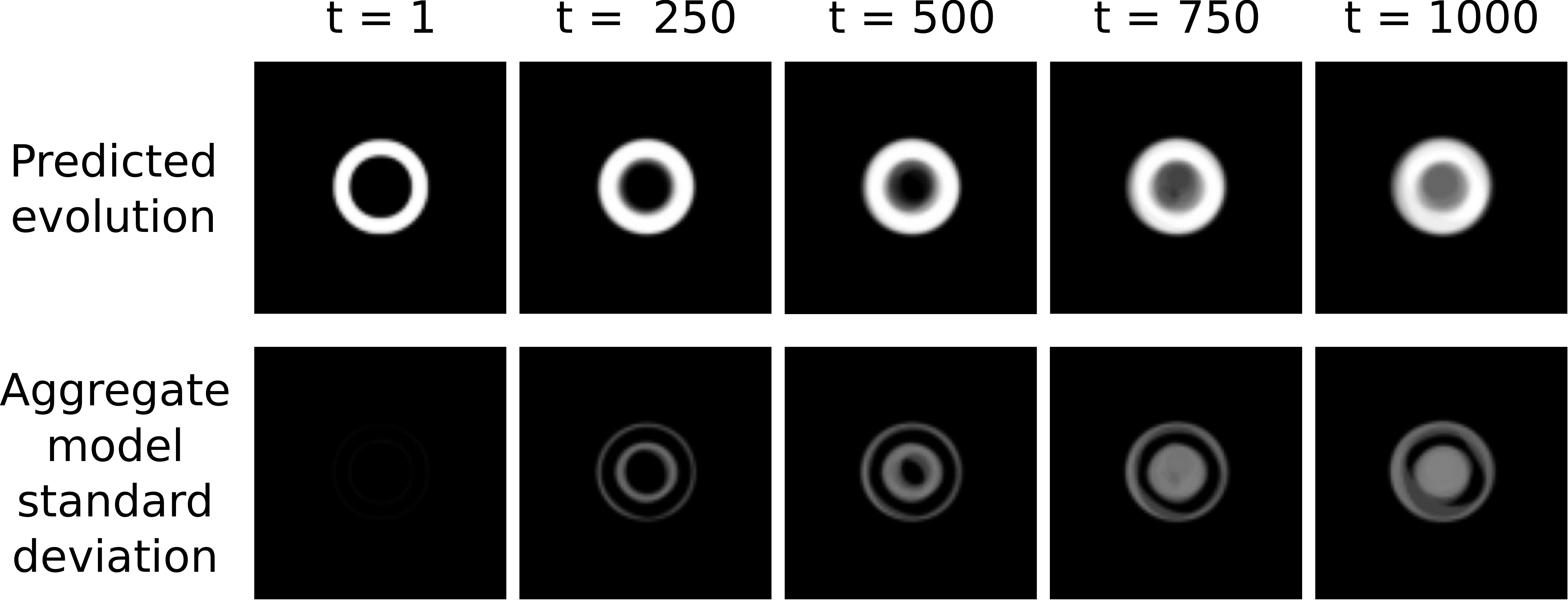}
    \caption{\label{fig::figure5} Evolution (top) and aggregate model standard deviation (bottom) of a stable annular configuration. This kind of topology was not present in the training set. As soon as the evolution trajectory deviates from the correct stable configuration, aggregate uncertainty increases.}
\end{figure}

The evolution and the corresponding aggregate model standard deviation is reported in Fig.~\ref{fig::figure5}(top) and Fig.~\ref{fig::figure5}(bottom) respectively. As it can be clearly observed, the annular configuration is stable for times corresponding to approximately 500 snapshots, more than 3 times the length of the training sequences. This confirms that the NN model is both effective in recognizing this unobserved steady state configurations and in generalizing to sequences longer than the ones provided during training.

After 500 snapshots, a progressive deviation of the evolution, caused by models in the aggregate unphysically filling the hole, can be observed. As soon as this happens, however, a clear increase in the prediction uncertainty is reported by the formation of the halo visible in the predicted $\varphi$ at t=750. Correspondingly, an increase of the standard deviation in the hole region can be observed, confirming the extrapolation recognition capabilities of our approach.

\subsection{Size dependent trends}
As reported in section~\ref{sec::train_valid_test}, training and validation shapes, as well as the initial test on Y configurations, all shared a common characteristic arm thickness. In order to check the ability of this ML method to replicate scale dependent phenomena, we decided to analyze to evolution of ellipsoidal domains. Self-similar shapes present a self-similar evolution under the (degenerate) Cahn-Hilliard equation~\cite{Herring1950JAP, Bray2002AiP, Andrews2020PRM}. Therefore, ellipsoidal domains will progressively reduce their aspect ratio (AR) starting from the initial value and reaching a circular equilibrium shape with a rate depending on the size of the initial condition.

\begin{figure}
    \includegraphics[width=\linewidth]{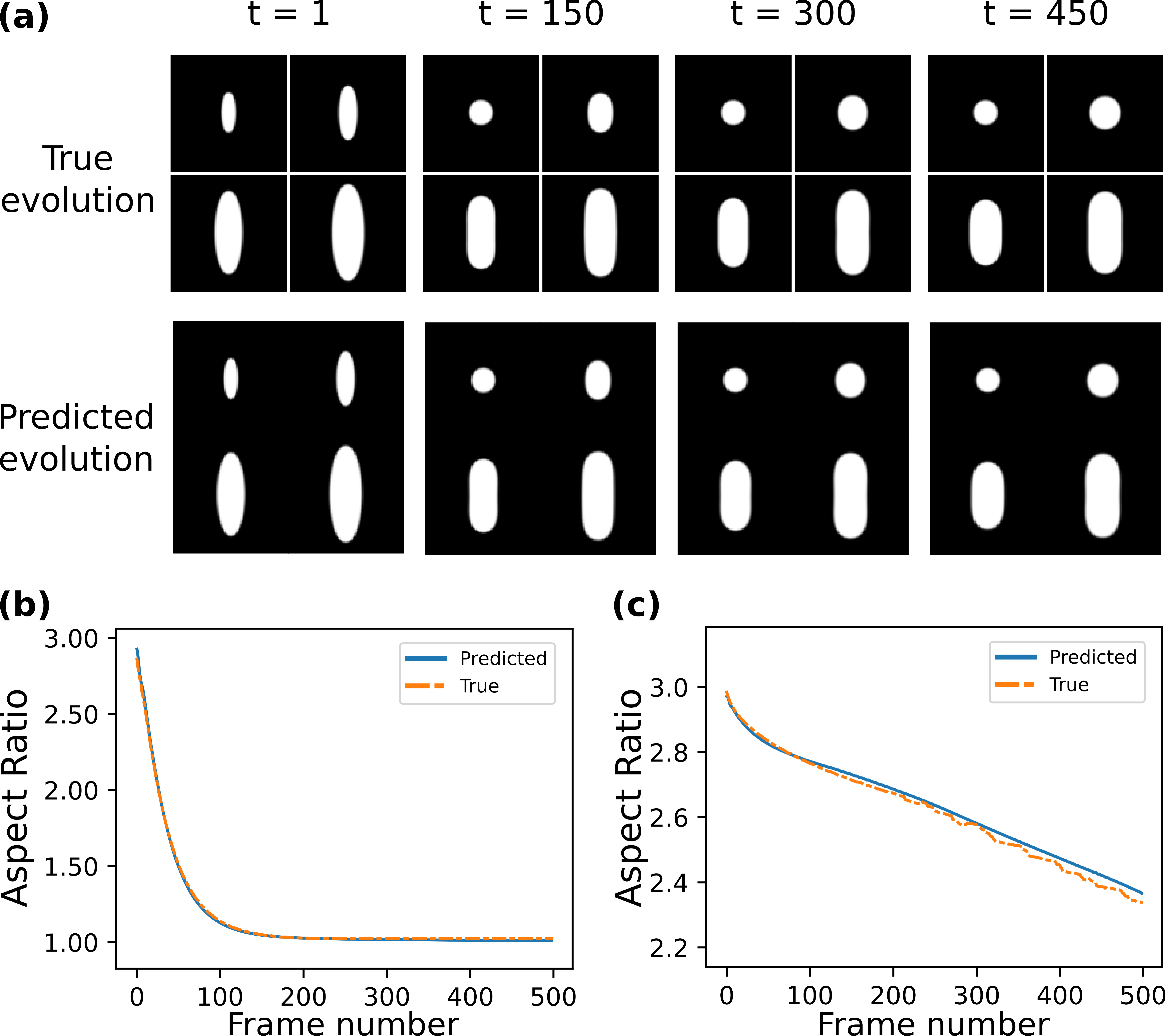}
    \caption{\label{fig::figure6} (a) Evolution of elliptical domains. The learned model has implicitly learned time scaling rules with the initial shape size. As computational costs depends on the size of the image for Convolutional Recurrent Neural Networks, all domains are evolved concurrently in the same simulation cell. (b) and (c) aspect ratio evolution for the smaller and bigger ellipses respectively as a function of frame number. Blue continuous line represents the AR predicted by the machine learning approach and dashed orange line reports the "True" aspect ratio as obtained by FEM solver.}
\end{figure}

To check if the CRNN was capable of capturing this trend, we compared the evolution of ellipses with an initial AR of 3 as simulated by FEM solver and by the ML approach. Fig.~\ref{fig::figure6}(a) reports such comparison. As can be observed, the NN evolution closely follows the high-fidelity phase field one. In Fig.~\ref{fig::figure6}(b) and (c), the aspect ratio as a function of time is also reported for the smallest and biggest ellipse, as a further proof of consistency. We emphasize again that the number of snapshots in these evolution is greater (more than 3 times) that of the sequences used during training.

The sequence of Fig.~\ref{fig::figure6} also highlights another characteristic of the NN approach. While PF evolutions of ellipses have been run independently, in the ML based prediction the evolution of all domains is calculated simultaneously by placing all of them in the same simulation cell. This does not increase computational costs for a cell of fixed size, as the number of operations for a convolutional neural network only depends on the number of pixels present in the image. This is in contrast to phase field approaches based on FEM, whose cost depends on the complexity of shapes due to mesh refinement. As all ellipses can be fitted in a single image, the overall computational cost is equivalent to treating ellipses independently. Additionally, this example shows explicitly how the Machine Learned evolution is both equivariant with respect to the group of spatial translations and local.

\subsection{Extension to complex conditions}

\begin{figure*}
    \includegraphics[width=\linewidth]{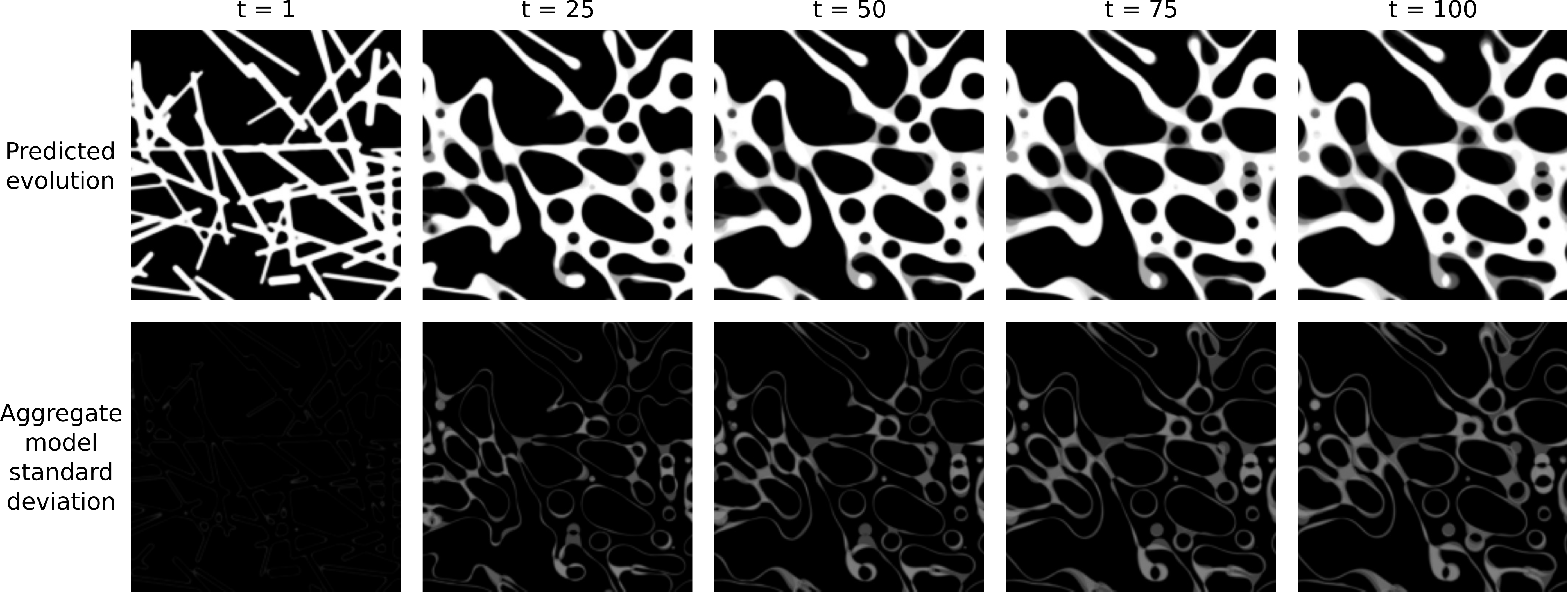}
    \caption{\label{fig::figure7} Evolution of complex initial conditions comprising many interacting domains. Critical topological changes such as pinching, which was not present in the training set, are marked as high prediction error areas.}
\end{figure*}

Finally, we show performances of our method in the case of complex intial conditions containing multiple interacting material domains and discuss limitations of our approach in these settings. We recall that the training set did not contain any evolution involving interacting domains, therefore these conditions represent an extrapolation test. Fig.~\ref{fig::figure7} shows the evolution of one of such configurations. Notice that an important computational effort would be required in order to treat the problem with traditional FEM solvers, as fine meshes are required to track the domain boundaries topology. On the other hand, NN evolution is orders of magnitude cheaper, as discussed in previous sections. As it is clear from Fig.\ref{fig::figure7}, the predicted evolution quickly exhibits a strong increase in prediction uncertainty in regions where non-trivial topological changes occur. This is made evident by the formation of halos and the increase of the variance between predictions of models in the ensemble.

These topological changes arise from well known behaviors in systems evolving by surface diffusion. In particular, high aspect ratio configurations undergo a Rayleigh-like instability and split into two or more domains as a result of a phenomenon known as pinch-off~\cite{Salvalaglio2018APL}. Notice that, even though such configurations were not present in the training set, the implemented prediction uncertainty estimation scheme is capable of identifying regions of the simulations where topological changes are likely to happen.

A less evident source of prediction uncertainty is present in the evolution of Fig.~\ref{fig::figure7}. Due to the existence of a lower bound in the representable physical dimension induced by the pixel discretization of the phase field, domains separated by less than such length are spuriously connected by gray regions. This can be noticed in Fig.\ref{fig::figure7} at t=1: when the distance between two domains is comparable or smaller than the pixel size, gray areas are formed due to resolution limits. This affects the subsequent evolution, as the NN model has no access to information on the high-resolution FEM counterpart. Regions containing close domains at t=1 also exhibit high uncertainty at subsequent times. While this phenomenon is still detected by the aggregate model standard deviation as a source of uncertainty, initial conditions for full NN evolutions should be considered carefully to avoid such pathological states. A possible approach, though not tested in depth in the current work, is to exploit the CRNN ability to process subsequences and feed the ML model more than a single state of the FEM evolution as an initial condition. The information lost by resolution downscaling may therefore be recaptured in temporal correlations between successive states of the system.

\section{Conclusions}
\label{sec::conclusions}

In this paper we have shown that CRNN can be conveniently used to learn morphological evolution via surface-diffusion. Even by training the model over a dataset including only extremely simple evolutions (simple shapes, single domains throughout the whole evolution), reliable extrapolation to both more complex shapes and long time scales has been proven. 

After introducing a convenient on-the-fly estimate of the uncertainty prediction, we showed that the model, instead, becomes less reliable during some critical evolution, involving e.g. lateral aggregation and splitting of a single domain via pinching. Limitations induced by the pixel-level discretization of the computational domains have also been discussed. Our results naturally open two further directions for future work: smart completion of the shape-evolution dataset by focusing solely on the high uncertainties regions, and designing a hybrid ML-PF approach where one automatically switches~\cite{MontesdeOcaZapiain2021npj} between the (computationally expensive) high-fidelity PF approach to the RCNN one based on the uncertainty prediciton. The latter would also remove pathological behaviors induced by finite resolution, as FEM solvers are capable to refine meshes as needed.

Finally, while here the methodology was tested under simple isotropic conditions and in the absence of external fields, suitable extensions to treat phenomena such as faceting~\cite{Salvalaglio2015CGD}, kinetically-limited growth~\cite{Albani2021SciRep} and/or heteroepitaxy~\cite{Rovaris2016PRB} are envisaged.

\begin{acknowledgments}
We wish to thank Prof. M. Ceriotti for urging us to add uncertainty predictions to our procedure. We acknowledge the CINECA award under the ISCRA initiative, for the availability of high-performance-computing resources and support.
\end{acknowledgments}

\end{document}